\documentclass[twocolumn]{aastex631}

\begin{document}

\title{Observations of Multiphase, High-Velocity, Shocked Gas in the Vela Supernova Remnant\footnote{Based on observations made with the NASA/ESA Hubble Space Telescope and the Far Ultraviolet Spectroscopic Explorer, obtained from the MAST data archive at the Space Telescope Science Institute, which is operated by the Association of Universities for Research in Astronomy, Inc., under NASA contract NAS5-26555. Based in part on data collected at the European Southern Observatory under ESO programs 194.C-0833(B) and 099.C-0637(A).}}

\correspondingauthor{Adam M.~Ritchey}

\author{Adam M.~Ritchey}
\affiliation{Eureka Scientific, 2452 Delmer Street, Suite 100, Oakland, CA 96402, USA}
\email{ritchey.astro@gmail.com}

\begin{abstract}
We present an analysis of high-resolution far-ultraviolet archival spectra obtained with the Space Telescope Imaging Spectrograph on the Hubble Space Telescope of the star HD~75309, which probes high-velocity shocked gas in the Vela supernova remnant (SNR). We examine high-velocity features from intrinsically strong absorption lines of O~{\sc i}, Si~{\sc ii}, Si~{\sc ii}*, C~{\sc ii}, C~{\sc ii}*, and Si~{\sc iii}. We also detect high-velocity components in the N~{\sc v} doublet and compare these features to observations of high-velocity O~{\sc vi} absorption, available from archival Far Ultraviolet Spectroscopic Explorer data. Kinetic temperatures are derived from the observed fractional abundances of the various ions, while gas densities and thermal pressures are obtained from the relative populations in excited fine-structure levels of C~{\sc ii} and Si~{\sc ii}. Our results indicate that the highly ionized species at high velocity probe gas in a region immediately behind a shock driven into an interstellar cloud, while the lower ionization species trace material further downstream in the cooling region of the post-shock flow. Low velocity N~{\sc v} and O~{\sc vi} absorption may trace gas in a conductive boundary layer between the unshocked portion of the cloud and the hot X-ray emitting intercloud medium. Temporal variations in high velocity Ca~{\sc ii} absorption features observed toward HD~75309 further confirm the highly inhomogeneous nature of the interstellar medium interacting with the Vela SNR.
\end{abstract}


\section{INTRODUCTION}
Numerous interstellar sight lines through the Vela supernova remnant (SNR) exhibit remarkable characteristics, including high velocity gas at high pressure \citep[e.g.,][]{jw95,j98} and temporal changes in absorption components at low and high velocity \citep[e.g.,][]{cs00,rao16,rao17,rao20}. Previous UV and visible absorption-line studies of stars probing the Vela SNR \citep[e.g.,][]{j76,j84,ds95,cs00} present a compelling picture of supernova-driven shocks interacting with a highly inhomogeneous interstellar medium (ISM). UV absorption-line spectroscopy is a powerful tool for examining the gas densities, kinetic temperatures, thermal pressures, and ionization states of interstellar clouds that have been shocked and accelerated by SNRs \citep[e.g.,][]{r20}. However, while there are numerous bright, early-type stars in the Vela region that could serve as suitable background targets for high-resolution UV spectroscopy, relatively few Vela targets have been observed using the high resolution echelle modes of the Space Telescope Imaging Spectrograph (STIS) onboard the Hubble Space Telescope (HST).

In this investigation, we examine archival HST/STIS spectra of the B1 IIp star HD~75309, located in the southeastern portion of the Vela SNR. High-resolution ground-based observations of HD~75309 reveal the presence of interstellar Ca~{\sc ii} absorption components at both high positive and high negative velocity \citep{cs00}. Moreover, the high velocity components exhibit changes in both equivalent width and velocity as a function of time \citep{cs00,p12}. \citet{cs00} found that a Ca~{\sc ii} component near $v_{\rm LSR}\approx-119$~km~s$^{-1}$ increased in equivalent width by 25\% between two observations obtained in 1993 and 1996. A second group of components at $v_{\rm LSR}=+81$~km~s$^{-1}$ and $+89$~km~s$^{-1}$ that were present in 1993 were completely absent from the spectrum taken in 1996. \citet{p12} obtained additional ground-based observations of HD~75309 in 2008. They reanalyzed the \citet{cs00} data, along with their new observations, finding that the high negative velocity component showed a systematic increase in velocity from $v_{\rm LSR}=-122$~km~s$^{-1}$ in 1993 to $-127$~km~s$^{-1}$ in 2008.\footnote{The velocities given in \citet{p12} are 4~km~s$^{-1}$ lower than those in \citet{cs00} for all components. The discrepancy seems to be related to differences in the corrections applied to place the spectra in the reference frame of the local standard of rest (LSR), combined with wavelength calibration uncertainties.} \citet{p12} also found that a high positive velocity Ca~{\sc ii} component toward HD~75309 increased in velocity from $v_{\rm LSR}=+119$~km~s$^{-1}$ in 1993 to $+124$~km~s$^{-1}$ in 2008.

\begin{figure*}
\centering
\includegraphics[width=0.67\textwidth]{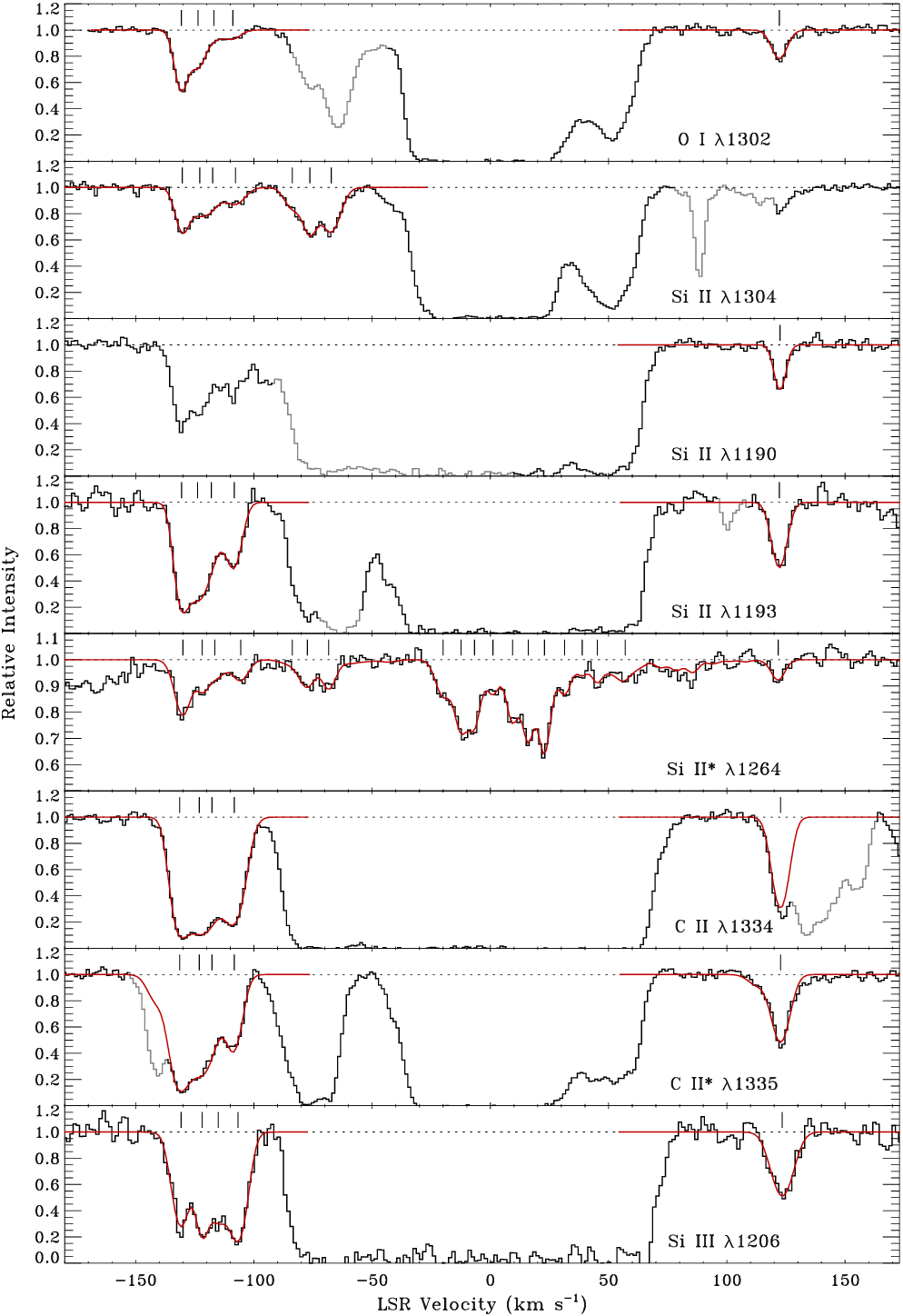}
\caption{Absorption profiles from HST/STIS observations of various atomic species that probe high-velocity gas toward HD~75309. The smooth red curves represent profile fits to specific absorption features where such fits are possible. Tick marks indicate the positions of the velocity components included in the fits. Gray portions of the histograms denote spectral regions where the absorption is blended with a transition from a different ion or fine-structure level (see the text for details).\label{fig:profiles}}
\end{figure*}

\citet{p12} interpreted the systematic velocity changes in the high positive and high negative velocity Ca~{\sc ii} components toward HD~75309 as evidence of the acceleration of interstellar clouds overtaken by the supernova blast wave. They used the measured accelerations ($\sim$10$^{-3}$~cm~s$^{-2}$) to model the cloud/shock interactions, finding total hydrogen column densities of $\sim$$6\times10^{17}$~cm$^{-2}$ for the accelerated clouds. In the context of a comprehensive analysis of O~{\sc vi} absorption in the Galactic disk, \citet{b08} describe an unusual O~{\sc vi} absorption profile toward HD~75309. They find that this sight line exhibits rather weak O~{\sc vi} absorption at $v_{\rm LSR}=+17$~km~s$^{-1}$, and much stronger absorption at $-90$~km~s$^{-1}$.\footnote{Here, we have applied a heliocentric-to-LSR velocity correction of $-16.3$~km~s$^{-1}$ to the measurements of \citet{b08} for the line of sight to HD~75309.} However, while \citet{b08} note the presence of unusually strong O~{\sc vi} absorption at high velocity toward HD~75309, no detailed analysis of this absorption is undertaken.

High-resolution HST/STIS observations of HD~75309 were obtained in 2000 as part of a SNAPSHOT survey of interstellar absorption lines (SNAP 8241; PI: J.~Lauroesch). However, while the moderate strength lines toward HD~75309 that trace high column density gas at low velocity (e.g., O~{\sc i}~$\lambda1355$, Mg~{\sc ii}~$\lambda\lambda1239,1240$, Mn~{\sc ii}~$\lambda1197$, Ni~{\sc ii}~$\lambda1317$, Cu~{\sc ii}~$\lambda1358$, Ge~{\sc ii}~$\lambda1237$, and Kr~{\sc i}~$\lambda1235$, along with numerous C~{\sc i} lines) have been included in various surveys of interstellar lines \citep{c01,a03,c06,jt11,j19}, the intrinsically strong absorption lines that trace low column density gas at high velocity (e.g., C~{\sc ii}~$\lambda1334$, O~{\sc i}~$\lambda1302$, and Si~{\sc ii}~$\lambda1304$) have yet to be analyzed in detail.

Here, we analyze the high velocity absorption features that appear in intrinsically strong lines toward HD~75309. These data allow us to derive estimates for the physical conditions in the high velocity shocked clouds during the time period when the clouds were observed to be accelerating \citep{p12}. We also analyze the absorption profiles of the highly-ionized species N~{\sc v} and O~{\sc vi} toward HD~75309. We discuss the implications of our results for models of SNRs interacting with a cloudy ISM.

\section{ARCHIVAL OBSERVATIONS}
Observations of HD~75309 were acquired on 2000 March 28 using the STIS/E140H grating, the $0\farcs2\times0\farcs2$ slit, and the central wavelength setting at 1271~\AA{}. This setting provides continuous wavelength coverage from 1160~\AA{} to 1356~\AA{}. An exposure time of 720 s yielded a signal-to-noise ratio (S/N) of $\sim$40 (per pixel) near 1264~\AA{}. The pipeline-processed spectra were retrieved from the Mikulski Archive for Space Telescopes (MAST). The individual echelle orders were merged into a single spectrum, weighting the flux values in the overlapping portions of the orders by the inverse square of their associated uncertainties. Small segments surrounding interstellar lines of interest (typically 2--3~\AA{} wide) were cut from the merged spectrum and were normalized via low-order polynomial fits to regions free of interstellar absorption.

\begin{figure*}
\centering
\includegraphics[width=0.67\textwidth]{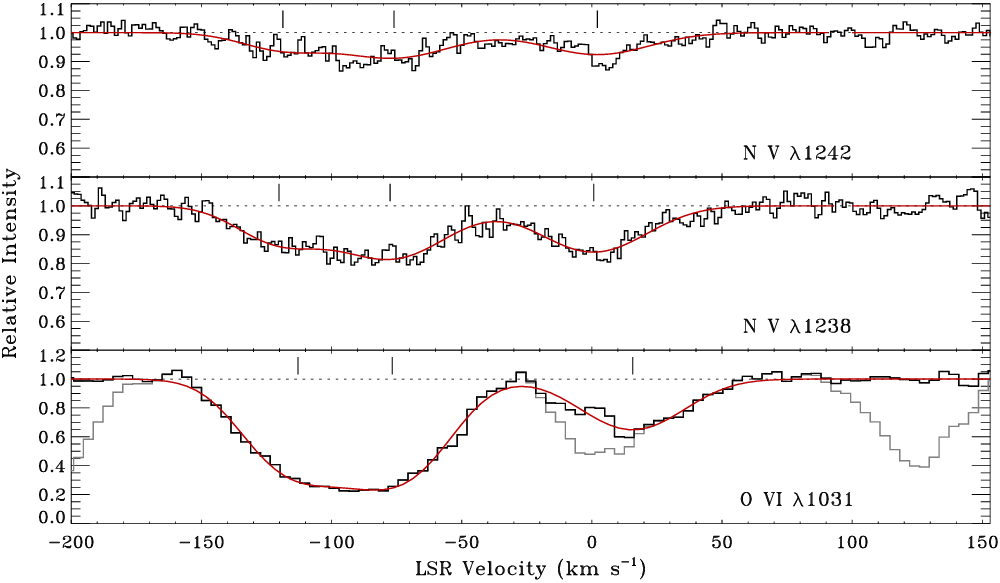}
\caption{Voigt profile fits to the N~{\sc v}~$\lambda\lambda1238,1242$ lines (from STIS spectra) and the O~{\sc vi}~$\lambda1031$ line (from FUSE data) toward HD~75309. The synthetic profiles are shown as smooth red curves, with histograms representing the observed spectra. Tick marks indicate the positions of the velocity components included in the fits. The gray histogram in the bottom panel shows the observed FUSE spectrum prior to removing the H$_2$ and HD lines in the vicinity of O~{\sc vi}~$\lambda1031$ (see the text for details).\label{fig:profiles2}}
\end{figure*}

Normalized absorption profiles for the atomic species showing high velocity absorption components toward HD~75309 are presented in Figure~\ref{fig:profiles}. In particular, we find high positive and high negative velocity components in O~{\sc i}, Si~{\sc ii}, Si~{\sc ii}*, C~{\sc ii}, C~{\sc ii}*, and Si~{\sc iii}.\footnote{In the usual spectroscopic notation, Si~{\sc ii} and C~{\sc ii} refer to the ground fine-structure levels of singly-ionized Si and C, while Si~{\sc ii}* and C~{\sc ii}* refer to the excited fine-structure levels.} The high positive velocity absorption appears as a single component in all species (with an average LSR velocity of $+122.5$~km~s$^{-1}$). The high negative velocity absorption shows a more complex component structure in the different ions, with individual component velocities ranging from $-108$~km~s$^{-1}$ to $-131$~km~s$^{-1}$. Nevertheless, the high positive and high negative velocity absorption seen in the UV lines appears to be closely related to the high positive and high negative velocity Ca~{\sc ii} components reported from ground-based observations \citep{cs00,p12}. An additional group of intermediate velocity components (with velocities ranging from $-68$~km~s$^{-1}$ to $-84$~km~s$^{-1}$) can be seen in the Si~{\sc ii}~$\lambda1304$ and Si~{\sc ii}*~$\lambda1264$ lines. These components are present in the other profiles shown in Figure~\ref{fig:profiles} as well but the absorption is either blended with another transition or is too strong to be reliably measured. No evidence of high or intermediate velocity absorption can be seen in the N~{\sc i} lines near 1200~\AA{} nor in the S~{\sc ii}~$\lambda\lambda1250,1253,1259$ triplet.

The STIS spectrum of HD~75309 shows absorption from the N~{\sc v}~$\lambda\lambda1238,1242$ doublet at both low velocity and high negative velocity (Figure~\ref{fig:profiles2}). However, no high positive velocity N~{\sc v} absorption can be seen. The N~{\sc v} absorption profile is therefore quite similar to the O~{\sc vi} profile discussed by \citet{b08}. For consistency, we reanalyzed the O~{\sc vi}~$\lambda1031$ line, which is available from observations obtained with the Far Ultraviolet Spectroscopic Explorer (FUSE). Fortunately, the FUSE observations of HD~75309 were obtained on 2000 January 26, very close in time to the STIS observations. Thus, any temporal changes in the absorption features between the two observation dates should be minimal.

The reduced FUSE spectra of HD~75309 were obtained from the MAST archive. For each detector segment, the eight individual exposures of HD~75309 were cross-correlated in wavelength space and then co-added by taking the weighted mean of the measured intensities. Ultimately, we used only the LiF1A and LiF2B detector segments for the O~{\sc vi} analysis since these had the highest S/N and showed the greatest consistency. The co-added spectra from these two detector segments were cross-correlated and co-added in the same manner as for the individual exposures. A small region surrounding the O~{\sc vi}~$\lambda1031$ line was cut from the data and the spectrum was normalized with a low-order polynomial. Our adopted continuum fit is similar to that shown in panel 79 of Figure~24 in \citet{b08}.

\section{PROFILE FITTING}
We derived column densities for the high velocity absorption components toward HD~75309 using the technique of multi-component Voigt profile fitting. The profile fitting routine, ISMOD \citep{s08}, treats the column densities, velocities, and $b$-values of the absorption components as free parameters while minimizing the rms of the fit residuals. A Gaussian instrumental line spread function is assumed, with $R=82,000$ for STIS E140H data obtained with the $0\farcs2\times0\farcs2$ slit \citep[see][]{s07}. (We adopt a FUSE resolving power of $R=17,000$.) In most cases, we are unable to fit the entire absorption profile because the absorption at low velocity is heavily saturated (see Figure~\ref{fig:profiles}). Instead, we fit only those absorption components that appear at high enough velocity that they are sufficiently isolated from the main saturated portion of the profile. An additional complication is that some of the absorption components are blended with other transitions. These cases are discussed in more detail below.

\begin{deluxetable*}{ccccccccccccc}
\tablecolumns{13}
\tablewidth{0pt}
\tabletypesize{\footnotesize}
\tablecaption{High Velocity Absorption Components toward HD~75309\label{tab:high_vel}}
\tablehead{ \colhead{$\langle v_{\rm LSR} \rangle$} & \colhead{$\log N$(O~{\sc i})} & \colhead{$b$(O~{\sc i})} & \colhead{$\log N$(C~{\sc ii})} & \colhead{$b$(C~{\sc ii})} & \colhead{$\log N$(C~{\sc ii}*)} & \colhead{$b$(C~{\sc ii}*)} & \colhead{$\log N$(Si~{\sc ii})} & \colhead{$b$(Si~{\sc ii})} & \colhead{$\log N$(Si~{\sc ii}*)} & \colhead{$b$(Si~{\sc ii}*)} & \colhead{$\log N$(Si~{\sc iii})} & \colhead{$b$(Si~{\sc iii})} }
\startdata
$-130.8$ & $13.41\pm0.03$ & 3.1 & $13.57\pm0.06$ & 4.1 & $13.61\pm0.06$ & 4.4 & $12.98\pm0.03$ & 3.4 & $11.76\pm0.04$ & 4.2 & $12.33\pm0.05$ & 4.2 \\
$-122.8$ & $13.16\pm0.03$ & 3.6 & $13.67\pm0.06$ & 6.6 & $13.45\pm0.05$ & 5.6 & $12.83\pm0.05$ & 5.3 & $11.21\pm0.10$ & 2.7 & $12.28\pm0.06$ & 2.7 \\
$-116.9$ & $12.57\pm0.11$ & 4.5 & $13.38\pm0.05$ & 7.5 & $13.19\pm0.04$ & 7.5 & $12.55\pm0.08$ & 4.8 & $11.33\pm0.12$ & 5.3 & $12.25\pm0.05$ & 4.3 \\
 $-107.6$ & $12.53\pm0.11$ & 4.1 & $13.53\pm0.06$ & 5.2 & $13.19\pm0.04$ & 3.9 & $12.65\pm0.07$ & 5.1 & $11.28\pm0.11$ & 4.0 & $12.47\pm0.06$ & 4.2 \\
 $-83.8$ & \ldots\tablenotemark{a} & \ldots\tablenotemark{a} & \ldots\tablenotemark{b} & \ldots\tablenotemark{b} & \ldots\tablenotemark{b} & \ldots\tablenotemark{b} & $12.65\pm0.05$ & 3.9 & $10.76\pm0.33$ & 5.5 & \ldots\tablenotemark{b} & \ldots\tablenotemark{b} \\
 $-76.9$ & \ldots\tablenotemark{a} & \ldots\tablenotemark{a} & \ldots\tablenotemark{b} & \ldots\tablenotemark{b} & \ldots\tablenotemark{b} & \ldots\tablenotemark{b} & $13.04\pm0.03$ & 3.6 & $11.38\pm0.09$ & 3.9 & \ldots\tablenotemark{b} & \ldots\tablenotemark{b} \\
 $-67.9$ & \ldots\tablenotemark{a} & \ldots\tablenotemark{a} & \ldots\tablenotemark{b} & \ldots\tablenotemark{b} & \ldots\tablenotemark{b} & \ldots\tablenotemark{b} & $13.12\pm0.03$ & 4.8 & $11.26\pm0.10$ & 3.2 & \ldots\tablenotemark{b} & \ldots\tablenotemark{b} \\
$+122.5$ & $13.08\pm0.04$ & 3.9 & $13.33\pm0.04$ & 4.0 & $13.21\pm0.03$ & 4.6 & $12.46\pm0.03$ & 3.1 & $11.18\pm0.11$ & 3.2 & $12.15\pm0.05$ & 5.7 \\
\enddata
\tablecomments{First column gives the average LSR velocity of the component. Units for $v$ and $b$ are km~s$^{-1}$. Units for $N$ are cm$^{-2}$.}
\tablenotetext{a}{Absorption from O~{\sc i}~$\lambda1302$ at this velocity is blended with low velocity absorption from P~{\sc ii}~$\lambda1301$.}
\tablenotetext{b}{Absorption component is too strongly saturated to yield a reliable measurement for $N$ or $b$.}
\end{deluxetable*}

We are able to fit the high positive and high negative velocity absorption in the O~{\sc i}~$\lambda1302$ line. However, the intermediate velocity absorption components are blended with low velocity absorption from P~{\sc ii}~$\lambda1301$. Similarly, we are able to fit the high negative velocity and intermediate velocity absorption in the Si~{\sc ii}~$\lambda1304$ line. However, in this case, the high positive velocity absorption is blended with low velocity absorption from O~{\sc i}*~$\lambda1304$\footnote{Absorption from the excited fine-structure levels of neutral oxygen (i.e., O~{\sc i}*~$\lambda1304$ and O~{\sc i}**~$\lambda1306$) is detected toward HD~75309, but only at low velocity (near $v_{\rm LSR}\approx1$ km~s$^{-1}$). This absorption appears to originate in normal, low-pressure interstellar material (A.~M.~Ritchey et al., in preparation).}. Fortunately, there are two other Si~{\sc ii} lines covered by the STIS observations, $\lambda1190$ and $\lambda1193$, both of which show unblended absorption from the high positive velocity component. The Si~{\sc ii}~$\lambda1193$ line also shows unblended absorption from the high negative velocity absorption complex. The high negative velocity components are seen in the Si~{\sc ii}~$\lambda1190$ line. However, in this case, the components are partially blended with a S~{\sc iii} transition at 1190.2~\AA{}.

The Si~{\sc ii}*~$\lambda1264$ line is weak enough that it is unsaturated at all velocities. However, the complication here is that the $\lambda1264$ line is part of a doublet. The weaker member of the doublet is shifted by +62.6~km~s$^{-1}$ and has an $f$-value that is $\sim$11\% of the stronger line \citep{m03}\footnote{All wavelengths and oscillator strengths for the transitions analyzed in this work are obtained from \citet{m03}.}. For our analysis, we fit both members of the Si~{\sc ii}* doublet simultaneously, keeping the velocities, $b$-values, and component column densities the same for the two transitions. Finally, the high positive velocity component in the C~{\sc ii}~$\lambda1334$ transition is blended with high negative velocity absorption from C~{\sc ii}*~$\lambda1335$. Furthermore, the C~{\sc ii}*~$\lambda1335$ line is actually a doublet, where the velocity separation between the two transitions is only 10.1~km~s$^{-1}$ but the weaker line again has an $f$-value that is $\sim$11\% of the stronger one. We therefore fit the C~{\sc ii} components at high positive and high negative velocity simultaneously with the same components in C~{\sc ii}* (avoiding the saturated portions of both profiles). For this fit, the velocities of the C~{\sc ii} and C~{\sc ii}* components were required to be the same but the $b$-values and column densities were allowed to vary between the two fine-structure levels.

The smooth red curves in Figure~\ref{fig:profiles} represent our profile synthesis fits to the various absorption components. The tick marks give the locations of the velocity components included in the fits. As mentioned above, the high positive velocity absorption is modeled with a single velocity component in all species. Whereas, four components are required to provide a satisfactory fit to the high negative velocity absorption complex. Table~\ref{tab:high_vel} provides the column densities, $b$-values, and average velocities of the components from the fits presented in Figure~\ref{fig:profiles}. Note that the low velocity components that are included in the fit to the Si~{\sc ii}* doublet (but are not included in any of the other fits) are excluded from Table~\ref{tab:high_vel}. The column density uncertainties reported in Table~\ref{tab:high_vel} account for the effects of noise in the spectra, uncertainties in continuum placement, and the degree of saturation in the absorption features. The velocities derived for corresponding components in different atomic species generally agree with one another. The typical standard deviation in velocity for corresponding components is 0.7 km~s$^{-1}$, which may be compared to the STIS velocity resolution of $\sim$3.7 km~s$^{-1}$.

\begin{deluxetable}{cccccc}
\tablecolumns{6}
\tablewidth{0pt}
\tabletypesize{\footnotesize}
\tablecaption{Highly Ionized Species toward HD~75309\label{tab:high_ions}}
\tablehead{ \colhead{$v$(N~{\sc v})} & \colhead{$\log N$(N~{\sc v})} & \colhead{$b$(N~{\sc v})} & \colhead{$v$(O~{\sc vi})} & \colhead{$\log N$(O~{\sc vi})} & \colhead{$b$(O~{\sc vi})} }
\startdata
$-119$ &  $12.99\pm0.07$ &  22 &  $-113$ &  $14.12\pm0.04$ &  22 \\
 $-77$ &  $13.30\pm0.05$ &  29 &  $-77$ &  $14.20\pm0.04$ &  22 \\
 $+1$ &  $13.21\pm0.05$ &  28 &  $+16$ &  $13.74\pm0.04$ &  24 \\
\enddata
\tablecomments{Units for $v$ and $b$ are km~s$^{-1}$. Units for $N$ are cm$^{-2}$.}
\end{deluxetable}

We synthesized the N~{\sc v} doublet toward HD~75309 with three components, one component at low velocity and two at high negative velocity (see Figure~\ref{fig:profiles2} and Table~\ref{tab:high_ions}). The stronger member of the doublet was fit first. Then, the derived $b$-values, component fractions, and relative velocities were held fixed in a fit to the weaker member. Before proceeding with the fit to the O~{\sc vi} absorption profile, two H$_2$ lines at 1031.2~\AA{} and 1032.4~\AA{} were modeled and removed from the spectrum. These lines do not impact the O~{\sc vi} absorption directly. A line of HD at 1031.9~\AA{}, however, does overlap with the O~{\sc vi} absorption profile. Following \citet{b08}, we created a model for the HD line, based on another HD line at 1021.5~\AA{}, which we then used to remove the HD absorption from the O~{\sc vi} profile. Finally, the O~{\sc vi} profile was fit with three velocity components for consistency with our analysis of the N~{\sc v} doublet. The parameters derived from our fits to the N~{\sc v} and O~{\sc vi} lines toward HD~75309 are provided in Table~\ref{tab:high_ions}.

\section{PHYSICAL CONDITIONS}
\subsection{Kinetic Temperatures}
The detection of high positive and high negative velocity absorption in multiple ions, and in both the ground and excited levels of Si~{\sc ii} and C~{\sc ii}, toward HD~75309 allows us to examine the physical conditions in these components in detail. \citet{gs07} have published theoretical predictions for the fractional abundances of different ions as a function of temperature for solar composition gas cooling radiatively at constant pressure. We can use these predictions to derive estimates for the kinetic temperatures and total hydrogen column densities of the shocked and accelerated clouds toward HD~75309.

\begin{figure}
\centering
\includegraphics[width=0.46\textwidth]{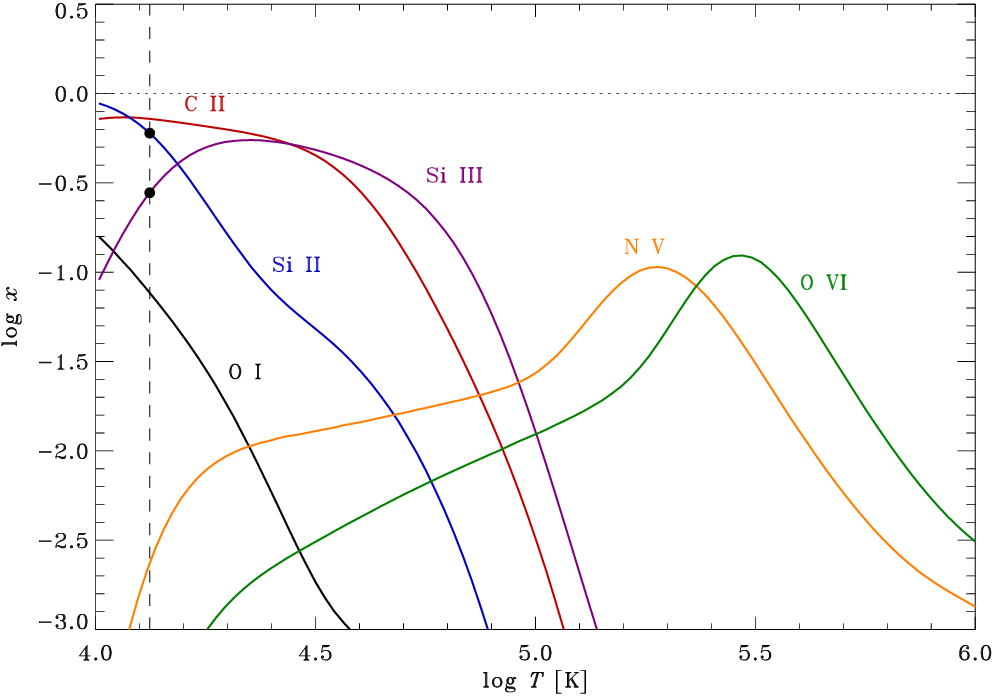}
\caption{Theoretical predictions for the fractional abundances of various ions as a function of temperature for solar composition gas cooling radiatively at constant pressure \citep{gs07}. The dots indicate the observed $N({\rm Si}~\textsc{ii}_{\rm tot})/N({\rm Si}~\textsc{iii})$ ratio for the +122.5~km~s$^{-1}$ component toward HD~75309. The vertical dashed line indicates the derived temperature based on this ratio.\label{fig:gs2007}}
\end{figure}

In Figure~\ref{fig:gs2007}, we plot the predicted fractional abundances\footnote{The fractional abundance of an ion is defined as (for example) $\log x({\rm O}~\textsc{i})=\log N({\rm O}~\textsc{i})-\log N({\rm H}_{\rm tot})-\log({\rm O}/{\rm H})_{\sun}$.} of O~{\sc i}, C~{\sc ii}, Si~{\sc ii}, Si~{\sc iii}, N~{\sc v}, and O~{\sc vi} as a function of temperature according to the models of \citet{gs07}. We adopt the nonequilibrium (isobaric) models because for $T\lesssim5\times10^6$~K the gas cools more rapidly than it can recombine \citep{gs07}. Since we have column density measurements for all of the ions shown we can use the ratios between different pairs of ions to estimate the kinetic temperatures. As an example, the observed $N({\rm Si}~\textsc{ii}_{\rm tot})/N({\rm Si}~\textsc{iii})$ ratio\footnote{Note that the total Si~{\sc ii} column density here is equal to $N($Si~{\sc ii}$_{\rm tot})=N($Si~{\sc ii}$)+N($Si~{\sc ii}*$)$.} of $\sim$2.2 for the $+122.5$~km~s$^{-1}$ component yields a kinetic temperature of $\sim$$1.3\times10^4$~K (Figure~\ref{fig:gs2007}). Other column density ratios for the $+122.5$~km~s$^{-1}$ component yield similar results. In particular, the $N({\rm C}~\textsc{ii}_{\rm tot})/N({\rm Si}~\textsc{iii})$, $N({\rm Si}~\textsc{iii})/N({\rm O}~\textsc{i})$, $N({\rm C}~\textsc{ii}_{\rm tot})/N({\rm O}~\textsc{i})$, and $N({\rm C}~\textsc{ii}_{\rm tot})/N({\rm Si}~\textsc{ii}_{\rm tot})$ ratios yield kinetic temperatures in the range $1.1$--$1.5\times10^4$~K. The standard deviation in these determinations for the $+122.5$~km~s$^{-1}$ component is $\sim$1600~K, indicating that the various determinations of kinetic temperature agree with one another at the $\sim$12\% level.

\begin{deluxetable*}{cccccccccc}
\tablecolumns{10}
\tablewidth{0pt}
\tabletypesize{\small}
\tablecaption{Derived Temperatures and Predicted Fractional Abundances (Low Ions)\label{tab:temp1}}
\tablehead{ \colhead{$\langle v_{\rm LSR} \rangle$} & \colhead{$\log \langle T \rangle$\tablenotemark{a}} & \colhead{$\log x$(H~{\sc i})} & \colhead{$\log x$(H~{\sc ii})} & \colhead{$\log x(e)$} & \colhead{$\log x$(O~{\sc i})} & \colhead{$\log x$(C~{\sc ii})} & \colhead{$\log x$(Si~{\sc ii})} & \colhead{$\log x$(Si~{\sc iii})} & \colhead{$\log \langle N \rangle$(H$_{\rm tot}$)\tablenotemark{b}} }
\startdata
$-130.8$ &  $4.06\pm0.02$ &  $-0.98$ &  $-0.048$ &  $+0.001$ &  $-0.93$ &  $-0.13$ &  $-0.11$ &  $-0.80$ &  $17.64\pm0.03$ \\
$-122.8$ &  $4.10\pm0.04$ &  $-1.10$ &  $-0.036$ &  $+0.015$ &  $-1.05$ &  $-0.14$ &  $-0.17$ &  $-0.63$ &  $17.52\pm0.08$ \\
$-116.9$ &  $4.16\pm0.05$ &  $-1.28$ &  $-0.023$ &  $+0.030$ &  $-1.23$ &  $-0.15$ &  $-0.32$ &  $-0.45$ &  $17.28\pm0.10$ \\
 $-107.6$ &  $4.19\pm0.06$ &  $-1.38$ &  $-0.019$ &  $+0.035$ &  $-1.32$ &  $-0.16$ &  $-0.40$ &  $-0.39$ &  $17.41\pm0.13$ \\
$+122.5$ &  $4.10\pm0.05$ &  $-1.11$ &  $-0.035$ &  $+0.016$ &  $-1.06$ &  $-0.14$ &  $-0.18$ &  $-0.62$ &  $17.32\pm0.12$ \\
\enddata
\tablenotetext{a}{Mean (and standard deviation) of the temperatures (in K) derived from the relative fractional abundances of different pairs of ions (see the text).}
\tablenotetext{b}{Mean (and standard deviation) of the hydrogen column densities (in cm$^{-2}$) implied by the predicted fractional abundances of the various ions and their observed column densities.}
\end{deluxetable*}

An implicit assumption in our use of the \citet{gs07} models to derive kinetic temperatures is that the fractional abundances of the different ions are not affected by depletion of the atoms onto dust grains. This is a valid assumption for high velocity clouds, where the dust grains are likely to have been destroyed through shock sputtering \citep[e.g.,][]{j94}. Nevertheless, we can test this assumption by examining the elemental abundance ratios for the high velocity components. Neither C nor O is likely to be depleted in the high velocity clouds toward HD~75309, since, even in low velocity diffuse molecular gas, the depletions of these elements are not very severe \citep{j09,r18}. However, Si is moderately depleted in normal quiescent gas \citep{j09}. The isobaric models of \citet{gs07} indicate that at $\sim$$1.5\times10^4$~K, most of the C will be in C~{\sc ii}, while the Si will be present as both Si~{\sc ii} and Si~{\sc iii}. We can therefore use the observed total C~{\sc ii} column densities and the sum of the total Si~{\sc ii} and Si~{\sc iii} column densities to obtain C/Si abundance ratios for the high velocity components. In this way, we find $\log ({\rm C}/{\rm Si})$ values in the range 0.81--0.93. These values are consistent (within the uncertainties which are $\sim$0.08~dex on average) with the solar ratio of $\log ({\rm C}/{\rm Si})_{\sun}=0.88$.\footnote{For consistency with the fractional abundances presented in \citet{gs07}, we adopt the solar abundances shown in their Table~1, which were obtained from \citet{a05}.}

Owing to the good agreement in the temperatures derived from the various ion pairs for the high velocity components toward HD~75309, the average temperatures are adopted for further analysis. The average kinetic temperatures derived in this way are presented in Table~\ref{tab:temp1}. The uncertainties listed for the temperatures correspond to the standard deviations of the results from the different ion pairs. We also list in Table~\ref{tab:temp1} the predicted fractional abundances of the ions based on the derived temperatures. These fractional abundances can be used in conjunction with the observed column densities to estimate the total hydrogen column density of each component. The total hydrogen column densities given in Table~\ref{tab:temp1} correspond to the mean values of the results implied by the predicted fractional abundances and observed column densities of O~{\sc i}, C~{\sc ii}, Si~{\sc ii}, and Si~{\sc iii}. The uncertainties in total hydrogen column density correspond to the standard deviations of these results. Note that each of the high velocity components is nearly completely ionized. We find hydrogen ionization fractions in the range $x({\rm H}^+)\approx0.89$--$0.96$. Thus, the total hydrogen column density is approximately equal to $N({\rm H}_{\rm tot})=N({\rm H}^0)+N({\rm H}^+)\approx N({\rm H}^+)$.

Additional constraints on the temperatures of the components can be obtained from the derived $b$-values. Since $b^2=(2kT)/m+2v_t^2$, where $T$ is the kinetic temperature, $k$ is Boltzmann's constant, $m$ is the atomic mass, and $v_t$ is the turbulent velocity, the measured $b$-value yields an upper limit on $T$ when $v_t$ is set equal to zero. In general, the $b$-values derived through profile fitting (Table~\ref{tab:high_vel}) are consistent with the temperatures given in Table~\ref{tab:temp1}. As an example, consider the $+122.5$~km~s$^{-1}$ component, which is isolated from other absorption components (except in the case of C~{\sc ii}) and therefore has fairly well-determined $b$-values. The maximum kinetic temperatures implied by the $b$-values derived from the O~{\sc i}, Si~{\sc ii}, Si~{\sc ii}*, C~{\sc ii}, and C~{\sc ii}* lines for this component are in the range $1.1$--$1.7\times10^4$~K. The larger $b$-value obtained from the Si~{\sc iii} line (5.7~km~s$^{-1}$) implies a maximum kinetic temperature of $5.5\times10^4$~K. This larger $b$-value could indicate that the Si~{\sc iii}-bearing gas has a somewhat broader distribution (or larger turbulent velocity) compared to the lower ionization species.

\begin{deluxetable}{ccccc}
\tablecolumns{5}
\tablewidth{0pt}
\tabletypesize{\small}
\tablecaption{Derived Temperatures (High Ions)\label{tab:temp2}}
\tablehead{ \colhead{$\langle v_{\rm LSR} \rangle$\tablenotemark{a}} & \colhead{$\log T$} & \colhead{$\log x$(N~{\sc v})} & \colhead{$\log x$(O~{\sc vi})} & \colhead{$\log N$(H$_{\rm tot}$)} }
\startdata
$-116$ &  $5.41\pm0.02$ &  $-1.21$ &  $-0.95$ &  $18.42\pm0.09$ \\
 $-77$ &  $5.38\pm0.01$ &  $-1.10$ &  $-1.04$ &  $18.62\pm0.06$ \\
 $+9$ &  $5.30\pm0.02$ &  $-0.98$ &  $-1.33$ &  $18.41\pm0.06$ \\
\enddata
\tablenotetext{a}{Average LSR velocity of the components from Table~\ref{tab:high_ions}.}
\end{deluxetable}

At kinetic temperatures of $1.1$--$1.5\times10^4$~K, as derived for the high negative velocity components seen in low ionization species toward HD~75309 (Table~\ref{tab:temp1}), the predicted fractional abundances of N~{\sc v} and O~{\sc vi} are very small (Figure~\ref{fig:gs2007}). This is in conflict with the rather large column densities of N~{\sc v} and O~{\sc vi} derived for the components at high negative velocity (Table~\ref{tab:high_ions}). The component at approximately $-116$~km~s$^{-1}$ in N~{\sc v} and O~{\sc vi} appears to be related to the group of components seen in the lower ionization species with velocities from $-108$~km~s$^{-1}$ to $-131$~km~s$^{-1}$. Similarly, the $-77$~km~s$^{-1}$ component in N~{\sc v} and O~{\sc vi} covers a similar velocity range as the Si~{\sc ii} components that appear at velocities from $-68$~km~s$^{-1}$ to $-84$~km~s$^{-1}$. However, the highly ionized species have much broader distributions, with $b$-values ranging from 22~km~s$^{-1}$ to 29~km~s$^{-1}$. Moreover, the column density of the O~{\sc vi} component at $-113$~km~s$^{-1}$ is \emph{larger} than the sum of the column densities of all of the O~{\sc i} components seen at high negative velocity. These observations therefore suggest the presence of \emph{multiphase} high-velocity shocked gas in the direction of HD~75309.

The $N({\rm O}~\textsc{vi})/N({\rm N}~\textsc{v})$ ratios for the three components included in the fits to these species yield kinetic temperatures in the range $2.0$--$2.6\times10^5$~K (Table~\ref{tab:temp2}). These temperatures are consistent with the $b$-values derived through profile fitting, which yield maximum kinetic temperatures in the range $4.1$--$7.0\times10^5$~K. The predicted fractional abundances of N~{\sc v} and O~{\sc vi}, combined with the observed column densities, yield total hydrogen column densities of $2.6$--$4.2\times10^{18}$~cm$^{-2}$. These column densities are considerably larger than the values associated with the high velocity clouds traced by the lower ionization species. For those lower ionization clouds, the total hydrogen column densities are in the range $1.9$--$4.3\times10^{17}$~cm$^{-2}$.

\subsection{Gas Densities and Thermal Pressures}
Now that we have established the kinetic temperatures of the high velocity clouds toward HD~75309, we can examine their gas densities and thermal pressures through an analysis of Si~{\sc ii} and C~{\sc ii} fine-structure excitation. Our analysis follows that of \citet{r20}, who examined similar observations for a line of sight through the SNR IC~443. Briefly, we consider the balance between collisional excitations to the upper fine-structure levels of Si~{\sc ii} and C~{\sc ii} and collisional and spontaneous de-excitations to the lower levels. Equations (3) and (7) in \citet{r20} present the equations for the Si~{\sc ii}*/Si~{\sc ii} and C~{\sc ii}*/C~{\sc ii} population ratios. These equations consider excitations by collisions with electrons, protons, and neutral hydrogen atoms, although electrons tend to dominate the collisional excitation of ions such as Si$^+$ and C$^+$ \citep[e.g.,][]{k85,k86}.

For Si~{\sc ii} excitation by electron impact, we adopt the rate coefficients of \citet{ak14}. Rate coefficients for C~{\sc ii} excitation by electron impact are derived from the expression given in \citet{g12}. \citet{b05} provide expressions for the rate coefficients for Si~{\sc ii} and C~{\sc ii} excitation by collisions with neutral hydrogen atoms. For excitations by collisions with free protons, we use the rate coefficients tabulated by \citet{bf70} for Si~{\sc ii} and by \citet{f96} for C~{\sc ii}. All of these rate coefficients are temperature dependent. We therefore use the derived kinetic temperatures of the high velocity components (Table~\ref{tab:temp1}) to obtain the appropriate rate coefficients in each case. The fractional abundances of H$^0$, H$^+$, and $e$ from Table~\ref{tab:temp1} are used to set the relative densities of these species for the collisional calculations.

\begin{figure}
\centering
\includegraphics[width=0.46\textwidth]{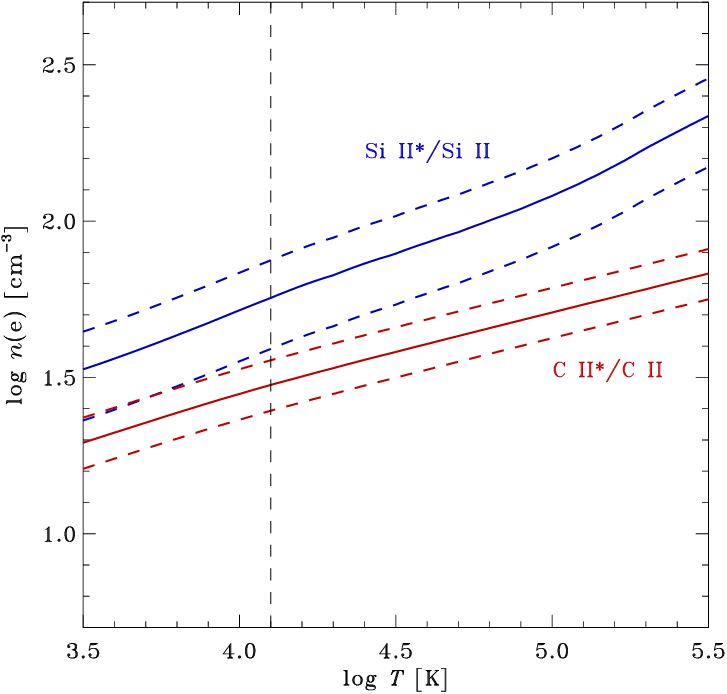}
\caption{Curves of the electron density versus temperature that correspond to the measured values of the ratios $N$(Si~{\sc ii}*)/$N$(Si~{\sc ii}) and $N$(C~{\sc ii}*)/$N$(C~{\sc ii}) for the $+122.5$~km~s$^{-1}$ component toward HD~75309. The solid curves correspond to the measured ratios, while the dashed curves give the 1$\sigma$ uncertainty ranges. The vertical dashed line represents the temperature of the +122.5~km~s$^{-1}$ component derived from various ion pairs (e.g., Figure~\ref{fig:gs2007}).\label{fig:c2_si2}}
\end{figure}

\begin{deluxetable*}{cccccccc}
\tablecolumns{8}
\tablewidth{0pt}
\tabletypesize{\small}
\tablecaption{Excitation Ratios and Derived Densities and Thermal Pressures (Low Ions)\label{tab:den}}
\tablehead{ \colhead{$\langle v_{\rm LSR} \rangle$} & \colhead{$N$(Si~{\sc ii}*)/$N$(Si~{\sc ii})} & \colhead{$\log n$(H$_{\rm tot}$)\tablenotemark{a}} & \colhead{$N$(C~{\sc ii}*)/$N$(C~{\sc ii})} & \colhead{$\log n$(H$_{\rm tot}$)\tablenotemark{b}} & \colhead{$\log \langle n \rangle$(H$_{\rm tot}$)\tablenotemark{c}} & \colhead{$\log (p/k)$\tablenotemark{d}} & \colhead{$\log L$\tablenotemark{e}} }
\startdata
$-130.8$ &  $0.060\pm0.008$ &  $1.80\pm0.05$ &  $1.10\pm0.23$ &  $1.77\pm0.18$ &  $1.80\pm0.05$ &  $6.18\pm0.05$ &  $-2.65\pm0.06$ \\
$-122.8$ &  $0.024\pm0.007$ &  $1.40\pm0.11$ &  $0.60\pm0.11$ &  $1.32\pm0.10$ &  $1.35\pm0.08$ &  $5.78\pm0.09$ &  $-2.32\pm0.11$ \\
$-116.9$ &  $0.061\pm0.022$ &  $1.82\pm0.14$ &  $0.65\pm0.10$ &  $1.37\pm0.09$ &  $1.40\pm0.08$ &  $5.89\pm0.09$ &  $-2.61\pm0.13$ \\
 $-107.6$ &  $0.042\pm0.014$ &  $1.66\pm0.12$ &  $0.45\pm0.08$ &  $1.15\pm0.09$ &  $1.19\pm0.08$ &  $5.72\pm0.10$ &  $-2.27\pm0.15$ \\
$+122.5$ &  $0.052\pm0.016$ &  $1.74\pm0.12$ &  $0.75\pm0.09$ &  $1.46\pm0.07$ &  $1.49\pm0.07$ &  $5.92\pm0.08$ &  $-2.66\pm0.14$ \\
\enddata
\tablenotetext{a}{Gas density (in cm$^{-3}$) derived from the observed $N$(Si~{\sc ii}*)/$N$(Si~{\sc ii}) ratio.}
\tablenotetext{a}{Gas density (in cm$^{-3}$) derived from the observed $N$(C~{\sc ii}*)/$N$(C~{\sc ii}) ratio.}
\tablenotetext{c}{Weighted mean value of the gas density.}
\tablenotetext{d}{Thermal pressure (in K cm$^{-3}$) derived from the mean density and the temperature given in Table~\ref{tab:temp1}.}
\tablenotetext{e}{Length scale (in pc) of the shocked component, defined as $L \equiv N({\rm H}_{\rm tot})/n({\rm H}_{\rm tot})$.}
\end{deluxetable*}

The total hydrogen densities, where $n({\rm H}_{\rm tot})=n({\rm H}^0)+n({\rm H}^+)\approx n({\rm H}^+)$, are then obtained from the measured $N$(Si~{\sc ii}*)/$N$(Si~{\sc ii}) and $N$(C~{\sc ii}*)/$N$(C~{\sc ii}) ratios. Strictly speaking, only one excitation ratio is required to derive the gas density. Having two ratios allows us to check for consistency between the two results. For each of the high velocity components toward HD~75309, the densities derived from the Si~{\sc ii} and C~{\sc ii} excitation ratios agree within their 2$\sigma$ mutual uncertainties. However, we find that the densities from the Si~{\sc ii} ratios tend to be systematically higher than those from the C~{\sc ii} ratios. An example is presented in Figure~\ref{fig:c2_si2}, where we plot curves of the electron density versus temperature that correspond to the measured values of the $N$(Si~{\sc ii}*)/$N$(Si~{\sc ii}) and $N$(C~{\sc ii}*)/$N$(C~{\sc ii}) ratios for the $+122.5$~km~s$^{-1}$ component. The vertical dashed line in the figure indicates the kinetic temperature of the component from Table~\ref{tab:temp1}. The two results for the density agree at the 1.4$\sigma$ level, but the curve that corresponds to the Si~{\sc ii} ratio is systematically higher than that for the C~{\sc ii} ratio. It could be that the two ions probe regions with slightly different densities or that there is some systematic offset in the rate coefficients that are used in the collisional calculations. Regardless, we find the agreement between the two density calculations satisfactory and adopt the weighted mean of the two results for each component.

The mean densities and associated thermal pressures of the high velocity components toward HD~75309 are presented in Table~\ref{tab:den}. The derived pressures exhibit only moderate variations from one component to the next, with a mean value in $\log (p/k)$ of 5.90. This is much larger than the mean thermal pressure derived for sight lines probing the quiescent ISM \citep[3.58;][]{jt11}, but is similar to the pressures derived for the high positive velocity clouds toward HD~43582, which probes shocked gas in IC~443 \citep{r20}.

The line of sight to HD~75309 was included in the survey of \citet{jt11}, who derived thermal pressures for the cold neutral medium from an analysis of C~{\sc i} fine-structure excitations. However, the C~{\sc i} lines toward HD~75309 are found only at low velocity. \citet{jt11} reported a weighted average value of $\log (p/k)$ toward HD~75309 of 3.41 for C~{\sc i} absorption in the range $-1.8\le v_{\rm LSR}\le +11.7$~km~s$^{-1}$. This range in velocity is very similar to the range expected for normal interstellar material participating in differential Galactic rotation. Based on the distance to HD~75309 \citep[$\sim$1800~pc;][]{bj21}, and the star's Galactic coordinates ($l=265\fdg86$; $b=-1\fdg90$), differential rotation would be expected to produce LSR velocities between $0$~km~s$^{-1}$ and $+12$~km~s$^{-1}$. It is likely, therefore, that the low velocity, low pressure, neutral gas toward HD~75309 represents quiescent line-of-sight material that is unrelated to the Vela SNR.

The last column of Table~\ref{tab:den} gives the length scales (or thicknesses) of the shocked, high velocity components toward HD~75309. The cloud thickness (or more precisely the line-of-sight component of the pathlength through the shocked region) is defined as $L \equiv N({\rm H}_{\rm tot})/n({\rm H}_{\rm tot})$. To calculate these quantities, we use the mean densities from Table~\ref{tab:den} along with the total hydrogen column densities from Table~\ref{tab:temp1}. These calculations yield length scales in the range $2.2$--$5.4\times10^{-3}$~pc for the high velocity components seen in the lower ionization species. In the next section, we discuss these results in more detail in the context of a model for shocked high velocity gas originally developed for the Vela SNR by \citet{j76}.

\section{DISCUSSION AND CONCLUSIONS}
The first detailed investigation of UV absorption lines associated with the Vela SNR was that of \citet{j76}, who studied Copernicus observations of two stars (HD~74455 and HD~75821) positioned behind the remnant. Their observations revealed high velocity clouds (at $-180$~km~s$^{-1}$ and $+90$~km~s$^{-1}$ toward HD~74455 and at $-90$~km~s$^{-1}$ toward HD~75821) in numerous low ionization species but also in more highly ionized species, such as N~{\sc v} and O~{\sc vi}. \citet{j76} proposed a model for cloud/shock interactions in the Vela SNR, in which the blast wave, propagating in a low density intercloud medium, drives secondary shocks into denser clouds overtaken by the primary shock wave. \citet{j76} identified the high velocity N~{\sc v} and O~{\sc vi} components with gas in a region immediately behind the cloud shock. Whereas, the lower ionization species (e.g., C~{\sc ii} and N~{\sc ii}) were assumed to arise in a cooling region far downstream from the shock front propagating through the cloud. The N~{\sc v} and O~{\sc vi} components observed at low velocity were interpreted as arising from a conductive boundary layer at the interface between the unshocked portion of the cloud and the shocked intercloud medium. In such a scenario, the high temperatures of the low velocity N~{\sc v} and O~{\sc vi} components result from thermal conductive heating by the ambient X-ray emitting gas.

A similar scenario appears to be consistent with our analysis of the gas toward HD~75309. In this interpretation, the high negative velocity components seen in low ionization species (O~{\sc i}, C~{\sc ii}, Si~{\sc ii}, and Si~{\sc iii}), which have temperatures in the range $1.1$--$1.5\times10^4$~K, probe gas in the cooling region of a shock driven into a cloud overtaken by the SN blast wave. The associated N~{\sc v} and O~{\sc vi} component at $-116$~km~s$^{-1}$, which has a temperature of $\sim$$2.6\times10^5$~K, probes gas closer to the region immediately behind the shock front. A similar scenario would explain the association between the N~{\sc v} and O~{\sc vi} component at $-77$~km~s$^{-1}$ and the group of intermediate velocity components that we are able to analyze only in the Si~{\sc ii} and Si~{\sc ii}* lines. The low velocity N~{\sc v} and O~{\sc vi} component at approximately $+9$~km~s$^{-1}$ might then be interpreted as the conductive boundary layer for one or both of these shocked clouds. However, we cannot rule out an origin for the low velocity N~{\sc v} and O~{\sc vi} component as unrelated line-of-sight material. The column density of O~{\sc vi} at low velocity toward HD~75309 is consistent with the column density expected for a sight line through the Galactic disk, based on the midplane density of O~{\sc vi} from \citet{b08} and a pathlength of $\sim$1800 pc.

When an interstellar cloud is overtaken by a supernova blast wave, the shock driven into the cloud will have a velocity that depends on the density contrast between the cloud and the intercloud medium. If $v_b$ is the blast wave velocity, $v_c$ is the velocity of the cloud shock, and $\rho_c$ and $\rho_0$ are the pre-shock densities in the cloud and intercloud regions, then $v_c\approx v_b(\rho_0/\rho_c)^{1/2}$ \citep{mc75,k94}. For a shock propagating through a purely atomic cloud (for which $\gamma=5/3$), the gas immediately behind the shock will have a velocity of $\sim$$\frac{3}{4}v_c$. The velocity of the gas further downstream in the cooling region ($v_{\rm obs}$) will approach $v_c$ according to the relation $v_{\rm obs}=v_c[1-T_{\rm obs}/(4T_c)]$ \citep{fk80}, where $T_{\rm obs}$ is the observed temperature of the cooling region and $T_c$ is the temperature of the gas immediately behind the cloud shock.

Since the shock velocity is related to the post-shock temperature according to $v_c=(16kT_c/3\mu)^{1/2}$, we can use our estimates of the observed temperatures and velocities of the shocked components toward HD~75309 (Table~\ref{tab:temp1}) to derive values for both $v_c$ and $T_c$. For gas at the distance of the Vela SNR \citep[$\sim$290~pc;][]{d03}, the expected systemic velocity (relative to the LSR) is approximately +2~km~s$^{-1}$. This is very similar to the velocity of the dominant absorption component toward HD~75309, as seen in low ionization species such as O~{\sc i} and Kr~{\sc i} \citep[e.g.,][]{c01}. The observed velocities of the shocked components (at high negative velocity) are therefore in the range $v_{\rm obs}=|v-v_{\rm sys}|=110$ to 133~km~s$^{-1}$. (These are lower limits to the true velocities if there are significant transverse components to the shocks.) From the above relations, we find values of $v_c$ in the range 112--134~km~s$^{-1}$ and $T_c$ in the range $2.0$--$2.9\times10^5$~K. \citep[Here, we have assumed that $\mu\approx0.7m_{\rm H}$, appropriate for a warm, partially ionized pre-shock medium;][]{sd17}. The post-shock temperatures calculated in this way are in very good agreement with the temperature derived for the N~{\sc v} and O~{\sc vi} component at $-116$~km~s$^{-1}$ (Table~\ref{tab:temp2}). This tends to corroborate our interpretation of the high velocity N~{\sc v} and O~{\sc vi} absorption as probing gas in an extended region immediately behind the cloud shock.

\begin{figure*}
\centering
\includegraphics[width=0.67\textwidth]{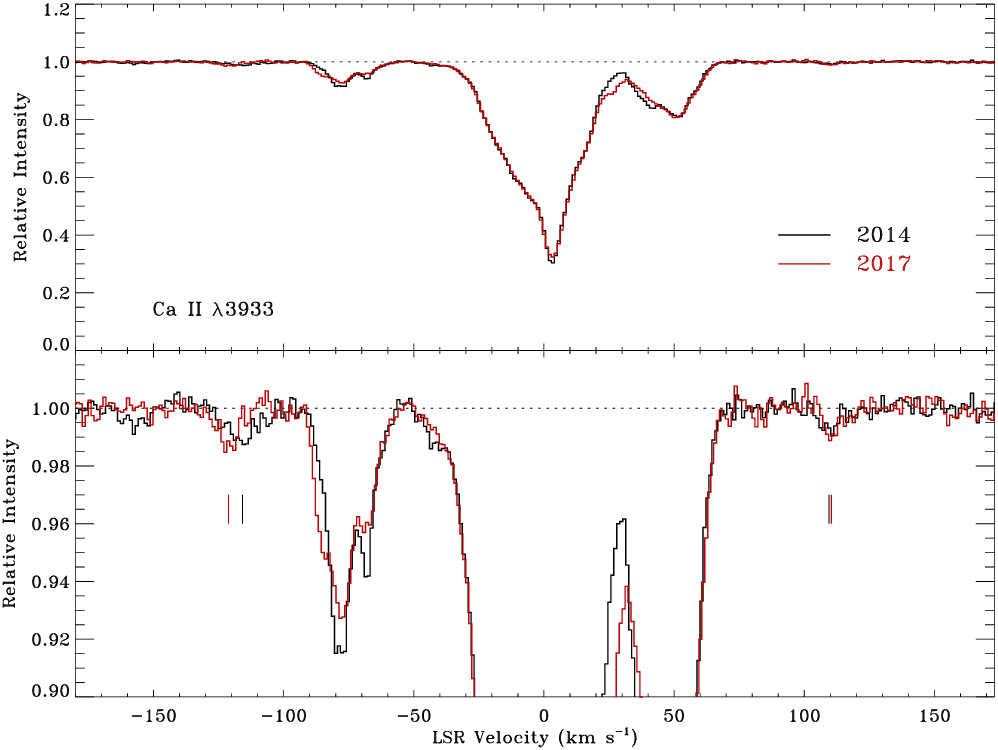}
\caption{High-resolution VLT/UVES spectra of HD~75309 showing absorption profiles of the Ca~{\sc ii}~K~$\lambda3933$ line. The spectrum obtained in 2014 is shown in black, while the 2017 spectrum is shown in red. The two panels show the same spectra but at different scales in relative intensity. Some of the differences between the two absorption profiles may be due to a slight change in the resolving power of the telescope. In the lower panel, the velocities of two very weak absorption components at high positive and high negative velocity are indicated with tick marks (black: 2014; red: 2017).\label{fig:cak}}
\end{figure*}

If we assume that isobaric cooling applies, then the post-shock temperatures derived above, combined with the observed densities and temperatures of the shocked low ionization components, yield post-shock densities in the range 1.0--2.3~cm$^{-3}$. For strong shocks, the shock compression ratio is $\sim$4. Thus, the pre-shock cloud densities are in the range 0.26--0.58~cm$^{-3}$. The blast wave velocity can be estimated from X-ray observations. A recent analysis by \citet{s20}, who examined X-ray emission from the entire Vela SNR, indicates that a two-temperature thermal plasma model provides the best fit to the X-ray data. The cooler of the two components, which \citet{s20} found to be in collisional ionization equilibrium at $T=2.2\times10^6$~K, implies a blast wave velocity of $v_b\approx400$~km~s$^{-1}$. This result for $v_b$ is in very good agreement with a determination based on a detailed examination of thermal and non-thermal X-ray emission from the Vela SNR by \citet{m23}. Adopting $v_b=400$~km~s$^{-1}$, the pre-shock density in the intercloud medium is in the range $n_0\approx n_c(v_c/v_b)^2=0.02$--$0.06$~cm$^{-3}$. These values are similar to, but somewhat lower than, the value of $\sim$0.1~cm$^{-3}$ derived by \citet{j76}.

The ion column densities predicted by sophisticated shock models \citep[e.g.,][]{a08} are generally consistent with our observations. For example, O~{\sc vi} column densities in excess of $10^{14}$~cm$^{-2}$ are produced by shocks with velocities $v_s\gtrsim175$~km~s$^{-1}$ \citep[for pre-shock densities in the range 0.1--1.0~cm$^{-3}$;][]{a08}. Likewise, N~{\sc v} column densities of $\sim$$10^{13}$~cm$^{-2}$ are produced by shocks with $v_s\gtrsim150$~km~s$^{-1}$. The fact that these velocities are somewhat higher than the observed velocities of the shocked clouds toward HD~75309 may indicate that the direction of shock propagation is tilted with respect to the line of sight. The observed (total) Si~{\sc iii} column density for the group of high negative velocity components toward HD~75309 is consistent with the model predictions for a 175~km~s$^{-1}$ shock if the magnetic field strength is larger than $\sim$2--4~$\mu$G \citep{a08}. However, for the lower ionization species, these same shock models predict column densities that are higher than the observed column densities by more than 1.1~dex for C~{\sc ii} and Si~{\sc ii} and more than 1.8~dex for O~{\sc i}. This indicates that the shock propagating through the cloud at high negative velocity is \emph{incomplete} and that the column densities of the low ionization species will continue to increase as the cooling and recombination proceeds.

Simple geometric arguments are consistent with a \emph{true} cloud shock velocity as high as $\sim$170~km~s$^{-1}$ toward HD~75309. At a distance of 290 pc, the radius of the Vela SNR is $\sim$20~pc, while the (on-sky) separation between HD~75309 and the Vela pulsar is $\sim$12~pc. If the shocked high negative velocity cloud toward HD~75309 is positioned near the approaching edge of the SNR shell, then a straight line connecting the pulsar with the cloud would make an angle of $\sim$$38\degr$ with the line of sight. This would imply that the true cloud velocities are in the range $v_{\rm obs}/\cos \theta=139$ to 168~km~s$^{-1}$. Repeating the above analysis, we find values of $v_c$ in the range 141--170~km~s$^{-1}$ and $T_c$ in the range $2.7$--$3.9\times10^5$~K. \citep[In this case, we have adopted $\mu\approx0.6m_{\rm H}$ for a fully ionized pre-shock medium;][]{sd17}. These values for $T_c$ would then yield post-shock densities in the range 0.8--1.7~cm$^{-3}$ and pre-shock cloud densities in the range 0.19--0.43~cm$^{-3}$.

As mentioned previously, the HST/STIS and FUSE spectra of HD~75309 were obtained at epochs intermediate between those corresponding to the Ca~{\sc ii} observations analyzed by \citet{cs00} and \citet{p12}. Thus, the physical conditions we derive for the high positive and high negative velocity clouds pertain to the time period during which these clouds were observed to be accelerating \citep{p12}. \citet{cs00} speculate that the increase in the equivalent width of the high negative velocity Ca~{\sc ii} component toward HD~75309 is a result of either the ongoing compression of the gas in the post-shock cooling region or the liberation of Ca ions from dust grains due to shock sputtering. While both of these mechanisms are undoubtedly important, a more complete census of the observed temporal variations along the line of sight to HD~75309 indicates that the actual situation is somewhat more complex.

\begin{deluxetable}{ccccc}
\tablecolumns{5}
\tablewidth{0pt}
\tabletypesize{\small}
\tablecaption{High Velocity Ca~{\sc ii} Components\label{tab:cak}}
\tablehead{ \colhead{Year} & \colhead{$v_{\rm LSR}$} & \colhead{$W_{\lambda}$(K)} & \colhead{$v_{\rm LSR}$} & \colhead{$W_{\lambda}$(K)} \\
\colhead{} & \colhead{(km s$^{-1}$)} & \colhead{(m\AA{})} & \colhead{(km s$^{-1}$)} & \colhead{(m\AA{})} }
\startdata
2014 & $-115.8$ & $2.1\pm0.4$ & $+109.5$ & $0.8\pm0.2$ \\
2017 & $-121.2$ & $1.7\pm0.3$ & $+110.4$ & $0.9\pm0.2$ \\
\enddata
\end{deluxetable}

New high-resolution ground-based spectra of HD 75309 were obtained in 2014 and 2017 using the Ultraviolet and Visual Echelle Spectrograph (UVES) on the Very Large Telescope (VLT)\footnote{The VLT/UVES spectra of HD~75309 were obtained under programs 194.C-0833(B) (PI: N.~Cox) and 099.C-0637(A) (PI: M.-F.~Nieva). These data were downloaded from the European Southern Observatory (ESO) Science Archive Facility and reduced using the UVES pipeline software.}. The absorption profiles of the Ca~{\sc ii}~K~$\lambda3933$ line from these observations are presented in Figure~\ref{fig:cak}. At first glance, the high positive and high negative velocity absorption components toward HD~75309 discussed by \citet{cs00} and \citet{p12} appear to have vanished in these more recent observations (upper panel of Figure~\ref{fig:cak}). However, upon closer examination, the high velocity components are present, but with much smaller equivalent widths and at somewhat lower (absolute) velocities (lower panel of Figure~\ref{fig:cak}). Moreover, the negative velocity component appears to have increased in velocity between 2014 and 2017, while the positive velocity component appears not to have changed substantially. Basic measurements involving the weak high velocity Ca~{\sc ii}~K components seen in the recent UVES spectra of HD~75309 are presented in Table~\ref{tab:cak}. These measurements may be compared with those in Table~5 of \citet{p12}.

The picture that emerges is one where the post-shock flow of cooling, recombining material passing in front of HD~75309 is chaotic and patchy, rather than smooth and homogeneous. Shocked gas components are accelerated and may grow in strength before vanishing beyond the line of sight. These components may then be replaced by additional ones carried along within the flow of shocked material passing in front of the star. Carefully planned multi-epoch HST/STIS observations of HD~75309, with a higher cadence than that which characterizes the available ground-based observations, would help to elucidate the changing physical conditions in the post-shock flow. The results of such an investigation would then provide useful constraints for 3D hydrodynamic simulations of SNR shocks interacting with a cloudy ISM.

\begin{acknowledgments}
This research is based on observations made with the NASA/ESA Hubble Space Telescope and the Far Ultraviolet Spectroscopic Explorer. Observations were obtained from the MAST data archive at the Space Telescope Science Institute, which is operated by the Association of Universities for Research in Astronomy, Inc., under NASA contract NAS 5-26555. The specific observations analyzed are associated with the HST program SNAP 8241 and the FUSE program P102. These data can be accessed via the following DOI: \dataset[10.17909/cq37-w558]{https://doi.org/10.17909/cq37-w558}. Ed Jenkins provided the package of IDL routines used to merge the STIS echelle orders. The FUSE spectra were processed using Don Lindler's LTOOLS package. Our results are also based in part on observations collected at the European Southern Observatory under ESO programs 194.C-0833(B) and 099.C-0637(A).
\end{acknowledgments}

\facilities{HST(STIS), FUSE, VLT(UVES)}
\software{ISMOD \citep{s08}}

\end{document}